\documentclass[journal]{IEEEtran}

\pdfoutput=1

\usepackage{amsmath,bm,amssymb,amsfonts,amsthm}
\usepackage[linesnumbered, ruled, vlined]{algorithm2e}
\usepackage{graphicx}
\usepackage{xcolor}
\usepackage{relsize}
\usepackage{textcomp}
\usepackage{listings}

\usepackage{changepage}
\usepackage{hhline}
\usepackage{multirow}
\usepackage{nomencl}
\usepackage{mathtools}
\usepackage{commath}
\usepackage{booktabs}
\usepackage{subfiles}
\usepackage{tabularx}
\usepackage{lscape}
\usepackage{rotating}

\usepackage[normalem]{ulem}
\usepackage[utf8]{inputenc}
\usepackage[hyphens]{url}
\usepackage{lineno}
    \modulolinenumbers[5]
\usepackage[caption=false]{subfig}
\usepackage{enumitem}
    \setlist{nolistsep}
\usepackage{gensymb}
\usepackage[hidelinks]{hyperref}
\usepackage{makecell}

\usepackage{scalerel}
\usepackage{tikz}
\usetikzlibrary{svg.path}
\definecolor{orcidlogocol}{HTML}{A6CE39}
\tikzset{
  orcidlogo/.pic={
    \fill[orcidlogocol] svg{M256,128c0,70.7-57.3,128-128,128C57.3,256,0,198.7,0,128C0,57.3,57.3,0,128,0C198.7,0,256,57.3,256,128z};
    \fill[white] svg{M86.3,186.2H70.9V79.1h15.4v48.4V186.2z}
                 svg{M108.9,79.1h41.6c39.6,0,57,28.3,57,53.6c0,27.5-21.5,53.6-56.8,53.6h-41.8V79.1z M124.3,172.4h24.5c34.9,0,42.9-26.5,42.9-39.7c0-21.5-13.7-39.7-43.7-39.7h-23.7V172.4z}
                 svg{M88.7,56.8c0,5.5-4.5,10.1-10.1,10.1c-5.6,0-10.1-4.6-10.1-10.1c0-5.6,4.5-10.1,10.1-10.1C84.2,46.7,88.7,51.3,88.7,56.8z};
  }
}
\newcommand\orcidicon[1]{\href{https://orcid.org/#1}{\mbox{\scalerel*{
\begin{tikzpicture}[yscale=-1,transform shape]
\pic{orcidlogo};
\end{tikzpicture}
}{|}}}}

\usepackage[noadjust]{cite}

\begin{document}

\title{\huge Relaxation Based Modeling of GMD Induced Cascading Failures in PowerModelsGMD.jl}

\author{
    Adam~Mate $^{1}$\orcidicon{0000-0002-5628-6509},
    Arthur~K.~Barnes $^{1}$\orcidicon{0000-0001-9718-3197},
    Steven~K.~Morley $^{2}$\orcidicon{0000-0001-8520-0199},
    Jacob~A.~Friz-Trillo $^{1,3}$\orcidicon{0000-0002-4377-8803},
    Eduardo~Cotilla-Sanchez $^{4}$\orcidicon{0000-0002-3964-3260},
    and Se\'{a}n~P.~Blake $^{5}$\orcidicon{0000-0001-9042-3557}
    \vspace{-0.25in}

\thanks{Manuscript submitted:~Aug.~15,~2021. 
Current version: Oct.~7,~2021.
}

\thanks{$^{1}$ The authors are with the Advanced Network Science Initiative at Los Alamos National Laboratory, Los Alamos, NM 87545 USA. \\ Email: amate@lanl.gov, abarnes@lanl.gov.}

\thanks{$^{2}$ The author is with the Space Science and Applications Group at Los Alamos National Laboratory, Los Alamos, NM 87545 USA. \\ Email: smorley@lanl.gov.}

\thanks{$^{3}$ The author is with the Department of Electrical and Biomedical Engineering at the University of Vermont, Burlington, VT 05405 USA. \\ Email: jacob.friz-trillo@uvm.edu.}

\thanks{$^{4}$ The author is with the School of Electrical Engineering and Computer Science at Oregon State University, Corvallis, OR 97331 USA. \\ Email: ecs@oregonstate.edu.}

\thanks{$^{5}$ The author is with the Heliophysics Science Division at NASA Goddard Space Flight Center, Greenbelt, MD 20771 USA. \\ Email: sean.blake@nasa.gov.}

\thanks{Acknowledgement: This work was supported by the U.S. Department of Energy -- Office of Cybersecurity, Energy Security, and Emergency Response.}

\thanks{LA-UR-21-28015. Approved for public release; distribution is unlimited.}

}

\markboth{IEEE/PES 53rd North American Power Symposium, November~2021}{}

\maketitle


\begin{abstract}
A major risk of geomagnetic disturbances (GMDs) is cascading failure of electrical grids.
The modeling of GMD events and cascading outages in power systems is difficult, both independently and jointly, because of the many different mechanisms and physics involved.
This paper introduces a relaxation based modeling of GMD-induced cascading failures:~the dc approximation-based DCSIMSEP solver was adapted to simulate cascading as a result of GMDs, the full set of ac power flow equations were relaxed to guarantee optimality, and the reactive power losses were modeled while keeping the problem convex.
The developed algorithm was implemented in PowerModelsGMD.jl -- an open-source software specifically designed to model and analyze geomagnetic hazards -- and demonstrated to work on the RTS-GMLC-GIC-EAST synthetic test network.
\end{abstract}


\section{Introduction} \label{sec:introduction}


Geomagnetic disturbances (GMDs) pose a serious threat to the continuous and reliable operation of the United States electrical grid. These solar-driven incidents disrupt the Earth's magnetic field and induce low-frequency electric fields on the surface.
The strongest geoelectric field disturbances are driven by auroral current systems and as the strength of a geomagnetic storm increases these auroral currents are driven to lower latitudes, exposing more populated regions to geoelectric hazard \cite{nikitina2016, EPRI2019, EPRI2020_blake}.
Geomagnetically induced currents (GICs) -- quasi-dc currents produced by these fields -- appear in the conductive infrastructure and flow into the high-voltage network through the neutrals of transformers \cite{pirjola2000-gic}.
Depending on their intensity, GICs can adversely impact transmission networks and equipment: by causing half-cycle saturation in transformers, harmonics are induced that may lead to the misoperation of protective devices, causing the tripping of over-current relays; due to overheating and thermal degradation, large high-voltage transformers are prone to premature aging, lasting damage, or complete failure; increased reactive power consumption, caused by the circulating GICs in the system, may lead to the loss of reactive power support and voltage collapses; in the worst case, widespread infrastructure damage and tripping of transmission lines may result in cascading outages and extended power disruptions \cite{NERC2012-gmd, overbye2013-gic, shetye2015-gmd, barnes21-hiddenfailures}.

\begin{figure}[!htbp]
\centering
\includegraphics[width=0.35\textwidth]{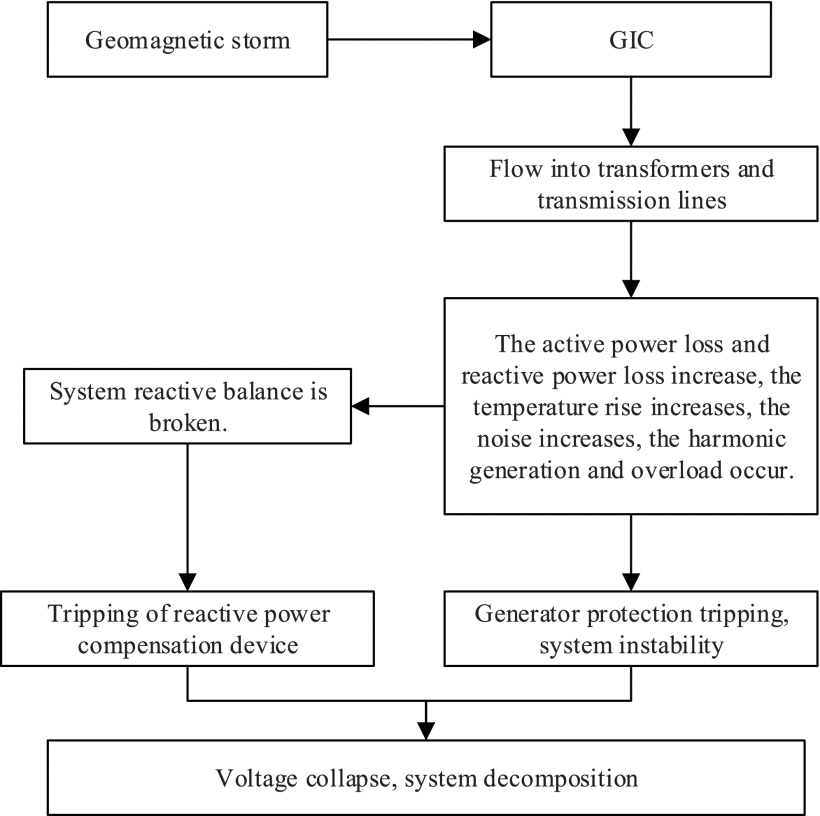}
\caption{Voltage collapse process due to a geomagnetic storm \cite{zhang-2020}.}
\label{fig:gmd-impact}
\end{figure}


The 2006 European blackout cascaded from a single 380~kV line in Germany, due to a mistimed disconnect, and within 28~seconds most of Europe's bulk power system had separated \cite{UTCE-2006}. A similar failure occurred in January 2021, but due to automatic load-shedding (ALS) and changes to the security criteria since 2006, the European grid narrowly escaped a continent-wide blackout \cite{EUPar-2021}.
Cascading failures occur faster than human operators can respond to, requiring a near instantaneous solution to prevent mass outages. Dynamic simulation and forensic analysis of cascading events suggest that the chain of events starts with a slow propagation regime, but usually transitions to a very fast regime of dependent failures \cite{song2016}. Full prevention of these failures is deemed near impossible due to the random nature in which such events occur and the vast number of cascading mechanisms involved in large-scale events.
A key answer to preventing transformer damage and system overload relies on the use of ALS and islanding of various zones when subsequent failures are likely. A variety of graph-based and machine learning approaches can be used as protection system support to decide which actions and when will be taken to prevent system collapse during a blossoming cascading failure \cite{meier2014-policy, ju2017, zhou2020-markovian}.

The expansion of renewable energy integration with an increasing demand on power grid resources could lead to a surge in cascading outages in the near future \cite{FERCKappenman-2010}. In addition, geomagnetic hazards will become more dangerous to energy infrastructures; with a solar cycle maximum occurring by 2026 and the probability of large geomagnetic storms increasing through 2029 \cite{owens2021}, the United States electrical grid could experience a cascading failure like the 1989 Hydro-Qu\'{e}bec GMD event on a modern, strained grid. Despite efforts in the past few decades to fortify the grid against GMDs and assure grid resilience -- e.g., improving network topology control, limiting transformer heating, deploying GIC blocking devices \cite{cicilio21-resilience} -- a major cascading event remains plausible.

The 1989 Hydro-Qu\'{e}bec event demonstrated the risk of GMDs causing cascading failures:~the province's electrical grid failed in 92~seconds, resulting in a blackout that left 6~million people without electricity for more than 9~hours \cite{DOE2019-gmd_monitoring, FERCKappenman-2010}.
The Qu\'{e}bec Interconnection is built on top of highly resistive rock, so when the geomagnetic storm struck, the 735~kV transmission lines experienced large GICs. Initially, a static VAR compensator (SVC) failed due to transformers experiencing half-cycle saturation and injecting excess reactive power into the system. Later, harmonics produced by GICs tripped protective relays covering SVCs, which then triggered the cascading failures of these devices throughout the grid. The system voltage rapidly collapsed as their regulation devices turned offline \cite{FERCKappenman-2010}.
Prior geomagnetic storm consequences -- e.g., strong voltage surges during the 1940 Philadelphia event, or damaged electrical equipment after the 1972 Canada event \cite{DOE2019-gmd_monitoring} -- and the aftermath of the Hydro-Qu\'{e}bec event led to preventive efforts, such as advanced protection methods and new modeling tools, aimed at mitigating similar future cascading failures across North America.


\vspace{0.1in}
The goal of GMD modeling is to realistically simulate GIC impact and determine threats that energy infrastructures face at any particular place and time in a power system.
As the level of detail required to model an electrical grid for this purpose is more than that needed for a traditional positive-sequence simulation, it is a complex task \cite{NERC2013-gic}. GICs are dependent on system characteristics (geographical location of substations, resistance of components, detailed transformer parameters), geomagnetic source fields (amplitude, frequency content, spatial characteristics), and the Earth conductivity structure (modeling method, substation grounding resistance, influence on geoelectric fields); all these need to be considered in the modeling process \cite{shetye2015-gmd, boteler2017-gic}.
PowerModelsGMD.jl\footnote{\url{https://github.com/lanl-ansi/PowerModelsGMD.jl}} (PMsGMD) \cite{PMsGMD} was developed in Julia \cite{software-julia}, a high-performance programming language for scientific computing, and was specifically designed to model and analyze geomagnetic hazards.

PMsGMD is an extension to PowerModels.jl \cite{software-PMs} and a member of the family of free and open-source packages under its umbrella that have been developed for simulating and optimizing infrastructure systems.
PowerModels.jl provides a platform to solve and evaluate steady-state power network optimization problems: it decouples problem specifications from the underlying problem formulations, allowing for convex relations to be easily applied.
Contrary to commercial software with GMD modeling tools -- e.g., PowerWorld\textsuperscript{\textregistered}, PSS\textsuperscript{\textregistered}E, MATLAB\textsuperscript{\textregistered} -- PMsGMD is an extensible open-source framework with verifiable and widely customizable capabilities with a focus on relaxations of power systems optimization problems. It allows for solving a variety of power systems optimization problems on a network subjected to GICs, such as power flow, optimal power flow (OPF), minimum load-shed (MLS), and optimal transmission switching (OTS).
The analysis of medium to large sized networks (a few hundred to thousands of buses) is quick and effortless. Industry standard format is used to define ac networks, however, the construction of related dc networks is required, which often is automatically generated in commercial software. Additionally, PMsGMD does not calculate geoelectric fields -- i.e., different Earth models are supported but not directly used, nor is coupling included -- it calculates GICs based on pre-determined geoelectric fields and takes in the coupled line voltages as inputs.


\vspace{0.1in}
While transformer overheating is a risk of GMD events, evidence suggests that such damage is sparse and can be prevented with appropriate thermal protective relaying that monitors the dc current in transformer neutrals \cite{dehghanian2021} or by treating overheating as an OTS problem \cite{PMsGMD}.
Cascading outages, due to the numerous failure points throughout the United States electrical grid \cite{barnes21-hiddenfailures}, are a greater concern and leave the grid vulnerable to blackouts.
In this paper, a GMD Cascade Simulator is introduced within the PMsGMD environment, which allows for modeling and analyzing GMD-induced cascading failures. To simulate cascading, DCSIMSEP \cite{eppstein2012} was adapted into the developed algorithm: the dc power flow was incorporated into the GIC power flow and relaxed ac power flow of PMsGMD. 
The result is the first ever model that captures both GMD and cascading failure together, which has the ability to approximate the dynamic behavior of power systems with high accuracy -- by using a relaxed MLS formulation that provides guaranteed convergence -- eliminating the need for complex dynamic or transient modeling.


\vspace{0.1in}
The remainder of this paper is organized as follows:
Section~\ref{sec:formulation} describes the modeling details and problem formulations that enable GMD-induced cascading failure modeling in PMsGMD. This includes the specifics of the analyzed GMD scenario and the induced geoelectric field calculations in Subsection~\ref{subsec:geoelectric-calc}, and the GIC coupling calculations that determine coupled line voltages for PMsGMD in Subsection~\ref{subsec:gic-coupling-calc}. Next, the developed GMD Cascade Simulator is presented in Subsection~\ref{subsec:cascading-failure-simulation}, followed by the descriptions of modeling formulations required for its algorithm: the transformer modeling formulations in Subsection~\ref{subsec:xfmr-modeling}, and the steady-state MLS and cascading MLS formulations in Subsection~\ref{subsec:steadystate-mls} and Subsection~\ref{subsec:cascading-mls}, respectively.
Section~\ref{sec:case-study} demonstrates the use of the Simulator on the created RTS-GMLC-GIC-EAST synthetic test network for the GMD scenario, and summarizes the conclusions of this work about the suitability of the Simulator for modeling GMD-induced cascading failures.
\newpage

\section{Problem Formulation} \label{sec:formulation}

\subsection{Geoelectric Field Calculations} \label{subsec:geoelectric-calc}

The GMD scenario is defined using a Space Weather Modeling Framework (SWMF) \cite{toth2012} simulation described by \cite{blake2021sims}, using the same model components used by the National Oceanic and Atmospheric Administration (NOAA) Space Weather Prediction Center for operations \cite{morley2018swmf}.
The performance of the operational configuration of SWMF at predicting GMDs has been demonstrated \cite{pulkkinen2013}, while improved predictions are obtained using higher spatial resolution grids to better resolve the current systems that ultimately drive the disturbance \cite{dimmock2021res}.
The specific scenario used here is the Scaled $A2$ scenario from \cite{blake2021sims} and represents a hypothetical scenario driven by solar wind observations between November~8 and 9, 2004 that were scaled up using physical considerations to obtain an extreme geomagnetic storm. This allows to define a temporally- and spatially-varying geoelectric field at any desired location, which cannot be achieved for historical events of this magnitude as they have limited observations and are challenging to reconstruct in detail \cite{wei2013, blake2021carr}.

The induced geoelectric field is related to the magnetic perturbation by the ground impedance, typically expressed as a frequency-dependent tensor $\mathbf{Z}(\xi)$, and often referred to as the magnetotelluric (MT) transfer function (TF). The MT TF is complex and typically expressed in practical units of [(mV/km)/nT].
The following is used to obtain the North and East components of the geoelectric field from the magnetic field perturbations:

\vspace{-0.1in}
\begin{subequations}
\allowdisplaybreaks
\small
\begin{align}
\begin{bmatrix} E_n(\xi) \\ E_e(\xi)
\end{bmatrix} =
\begin{bmatrix} Z_{nn}(\xi) & Z_{ne}(\xi) \\ Z_{en}(\xi) & Z_{ee}(\xi) 
\end{bmatrix}
\begin{bmatrix} B_n(\xi) \\ B_e(\xi)
\end{bmatrix} \\
E_n(t) = \mathcal{F}^{-1}(E_n(\xi)) \qquad
E_e(t) = \mathcal{F}^{-1}(E_e(\xi))
\end{align}
\label{eq:geoelec}
\end{subequations}
\noindent where $\xi$ denotes frequency and $B(\xi) = \mathcal{F}(B(t))$ using $\mathcal{F}$ to denote the Fourier transform; the subscripts $n$ and $e$ refer to geographic North and East, respectively.

In this work, $\mathbf{Z}(\xi)$ is obtained from the Incorporated Research Institutions for Seismology (IRIS) database of empirical MT TFs \cite{kelbert2018spud}, derived from MT surveys at defined sites across North America.
The case study in Section~\ref{sec:case-study} uses data from the USArray BB \cite{usarray_bb} and TA \cite{usarray_ta} surveys, so each $\mathbf{Z}(\xi)$ in the database corresponds to the MT TF of a specified location.

For each grid point on the surface of the Earth (at 1$^{\circ}$ resolution in latitude and longitude), the components of $\mathbf{B}(t)$ are taken, a cosine taper is applied to reduce spectral leakage \cite{pilz2012}, and fast Fourier transform is used to obtain $\mathbf{B}(\xi)$; the $\mathbf{Z}(\xi)$ is taken for the nearest survey site.
The complex-valued $\mathbf{Z}$ is interpolated to the required $\xi$ using a cubic spline interpolant in log($\xi$). 
Finally, $\mathbf{E}(t)$ is obtained by Eq.~\eqref{eq:geoelec}.

\begin{figure}[!htbp]
\centering
\includegraphics[width=0.475\textwidth]{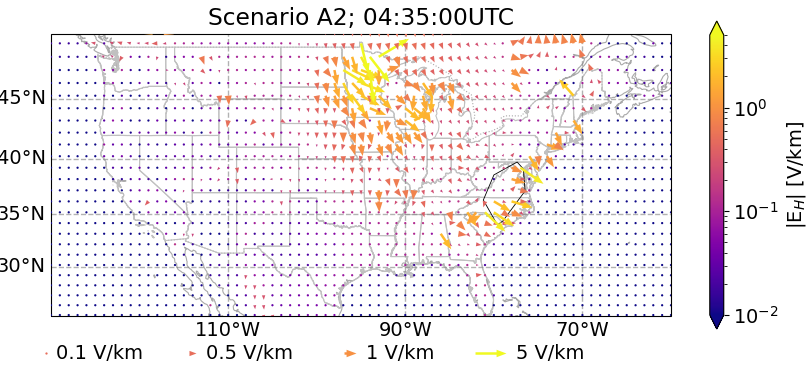}
\caption{
Horizontal geoelectric fields (E$_{H}$) over the continental United States at time 04:35:00 UTC in the $Scaled$ $A2$ scenario, when the largest values are reached over the RTS-GMLC-GIC-EAST convex hull. The convex hull is shown with the solid black line located near the U.S. East Coast.
}
\label{fig:conus}
\end{figure}

Figure~\ref{fig:conus} shows a snapshot of the predicted geoelectric fields for the $A2$ scenario \cite{blake2021sims}, taken at the time when geoelectric field magnitude peaks in the region defined by the convex hull of the RTS-GMLC-GIC-EAST network (described in Section~\ref{sec:case-study}).
The areas of strong geoelectric field are those that have both a strong geomagnetic driver and locally resistive geology. Therefore, severe geomagnetic storms such as this scenario can lead to peak geoelectric field magnitudes of $\sim$4\,V/km in the region covering Maryland to North Carolina.

\begin{figure}[!htbp]
\centering
\includegraphics[width=0.475\textwidth]{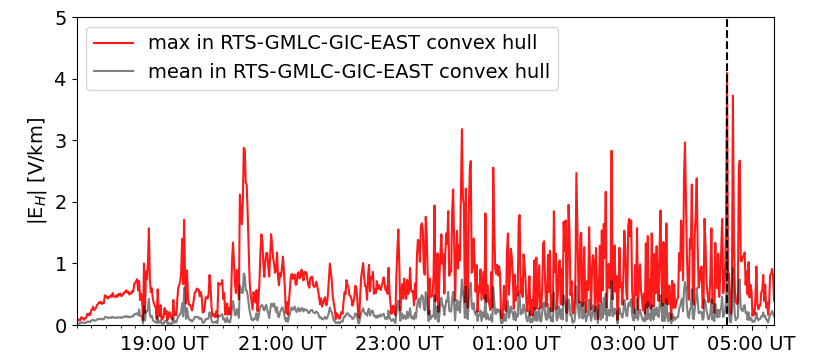}
\caption{
Geoelectric field magnitudes ($|$E$_{H}|$) in the region chosen for the case study. The red line shows the peak horizontal geoelectric field magnitude through the simulation, the black line shows the mean across the convex hull.
}
\label{fig:geoelec}
\end{figure}

Figure~\ref{fig:geoelec} shows the time series of the maximum and mean amplitudes of the predicted geoelectric fields for the $A2$ scenario in the region defined by the convex hull of the RTS-GMLC-GIC-EAST network. The vertical dashed line marks the time of the snapshot in Figure~\ref{fig:conus}.

\subsection{GIC Coupling Calculations} \label{subsec:gic-coupling-calc}

GIC coupling is calculated following the procedure of the North American Electric Reliability Corporation (NERC) \cite{NERC2013-gic}, which assumes a WGS84 Earth model.
The coupling model assumes that the variation of the geoelectric field takes place on a spatial scale that is large compared to the line length. Given a line with a sending endpoint $(x_i, y_i)$ and receiving endpoint $(x_j,y_j)$, the line midpoint is determined as:

\vspace{0.05in}
\noindent $ \forall ij \in \mathcal{E}^{d}$
\begin{equation}
\allowdisplaybreaks
\small
x_m = \frac{x_i + x_j}{2} \qquad
y_m = \frac{y_i + y_j}{2}
\label{eq:line-midpoint}
\end{equation}

\noindent and angular line displacement as:

\vspace{0.05in}
\noindent $ \forall ij \in \mathcal{E}^{d}$
\begin{equation}
\allowdisplaybreaks
\small
\delta_{ij}^x = x_j - x_i \qquad
\delta_{ij}^y = y_j - y_i
\label{eq:line-angular-displacement}
\end{equation}

\noindent Given the angular line displacement, the linear line displacement in units of [km] is calculated as:

\vspace{0.05in}
\noindent $ \forall ij \in \mathcal{E}^{d}$
\begin{subequations}
\allowdisplaybreaks
\small
\begin{align}
d_{ij}^e = 111.2 \cdot \delta_{ij}^y \cos(\gamma) \qquad
\gamma = \frac{\pi}{180}x_m \\
d_{ij}^n = 111.2 \cdot \delta_{ij}^x
\end{align}
\label{eq:line-linear-displacement}
\end{subequations}

\noindent Finally, the coupled line voltage is given as the inner product of the linear line displacement and the electric field vector at the measured point closest to the line midpoint:

\vspace{0.05in}
\noindent $ \forall ij \in \mathcal{E}^{d}$
\begin{equation}
\allowdisplaybreaks
\small
v^{dc}_{ij} = d_{ij}^e E_e + d_{ij}^n E_n
\label{eq:gic-coupling}
\end{equation}

\subsection{Cascading Failure Simulation} \label{subsec:cascading-failure-simulation}

The proposed cascade simulator -- formed by combining the problem formulations described in the next subsections -- is presented below:

\begin{algorithm}
\caption{GMD Cascade Simulator}
\label{alg:gmd-cascade}
\DontPrintSemicolon
\SetAlgoLined
\footnotesize
    run MLS with branch limits enabled and no induced voltage \;
    \While{true}{
        update generator setpoints and bounds \;
        update generator breakers \;
        update load breakers \;
        run MLS with branch limits enabled and induced voltage \;
        update branch overcurrent relays \;
        open branches with tripped relays \;
        \If{condition}{
            terminate \;
        }
    }
\end{algorithm}

The algorithm was designed to approximate the dynamic behavior of a power system under cascading conditions with a steady-state MLS formulation. This provides two major benefits:
First, the data requirements for steady-state formulations are relaxed compared to dynamic or transient formulations as information on components (e.g., governors, exciters, system stabilizers, dynamic load behavior, over/under-frequency relays, under-voltage relays) are not required; this allows for the use of many existing datasets for base-case power flow with minimal modifications.
Second, convex relaxations exist for the MLS formulation including the effect of GIC, which allows for continuing a time-series simulation in cases where a dynamic simulation may fail to converge and terminate.

\vspace{0.1in}
The GMD Cascade Simulator works by first performing a standard MLS at the initial time step, where generation is allowed to be dispatched (or disabled) independently, individual loads can be shed independently, and branch limits are respected. The algorithm then loops until either the simulation time has elapsed or the system has experienced a total blackout, indicated by all buses in the system being disabled.

At each iteration, the first step is to account for generator ramping by adjusting generator upper and lower bounds for active power based on the dispatched power from the previous iteration and maximum ramp rates; this assumes that generators have a fixed amount of time to ramp up or down.
The second step accounts for generator breaker opening by disabling any generators whose power output (accounting for continuous generator load-shed variables) is less than the generator lower bound for active power in the base case (not accounting for ramping).
The third step accounts for load breakers opening by setting the load powers based on the supplied load power at the previous iteration, ensuring that total load is monotonically decreasing over the course of the simulation.
The fourth step is to disable all components outside of those in the largest island (in terms of the number of buses). 
The fifth step is to update the generator ramping and generator/load breaker opening by running a modified MLS formulation; the generator and load-shed variables within the largest island are coupled together, emulating the effect of generator participation factors.
The sixth step is to update the internal integrator state for branch overcurrent relays.
Lastly, if the integrator state for any overcurrent relays exceeds the trip threshold, those branches are removed from service.
\vspace{0.05in}





\subsection{Transformer Modeling} \label{subsec:xfmr-modeling}

The formulation employed in PMsGMD contains two equivalent models of the power system: one for computing the ac power flows and one for computing the GICs. The main difference between them occurs in transformer (xfmr) modeling.
AC power flow models typically model transformers as a single edge with a voltage transformation (phase shift and tap change). GIC models, on the other hand, require models of transformers that include details of the series and common windings, as well as other transformer components \cite{zheng2014effects}; these are linked via reactive power losses.
Reactive power losses for transformer $ij$ are a function of the ``effective'' GIC $I_{ij}^{eff}$, which is computed using the following set of equations:

\vspace{0.05in}
\noindent $ \forall ij \in E^a $
\begin{equation}
\label{eq:effective_gic}
\allowdisplaybreaks
\small
I_{ij}^{eff} =
\begin{cases}
\abs{ I_{ij}^H } & \text{if $e \in E^\Delta$} \vspace{.1cm} \\
\abs{ I_{ij}^H + \frac{I_{ij}^L}{\alpha_{ij}} } & \text{if $ij \in E^y$} \vspace{.1cm}  \\
\abs{ \frac{\alpha_{ij} I_{ij}^S + I_{ij}^C}{\alpha_{ij} + 1} } & \text{if $ij \in E^\infty$} \vspace{.1cm} \\
\abs{ I_{ij}^H + \frac{I_{ij}^L}{\alpha_{ij}} + \frac{I_{ij}^T}{\beta_{ij}} } & \text{if $ij \in E^{3w}$} \vspace{.1cm} \\
0 & \text{otherwise}
\end{cases}
\end{equation}
\noindent where 1) models a gwye-delta generator step-up (GSU) xfmr, 2) models a gwye-gwye xfmr, 3) models a gwye-gwye auto-xfmr, and 4) models a three-winding gwye xfmr.

$E^a$ denotes the set of edges in the network corresponding to xfmrs: $E^\Delta$ is the set of delta-delta xfmrs, $E^y$, is the set of gwye-gwye xfmrs, $E^\infty$ is the set of gwye-wye auto-xfmrs, and $E^{3w}$ is the set of three-winding gwye xfmrs.
$I_{ij}^H$ is the GIC flowing through the xfmr primary winding and $I_{ij}^L$ is through the xfmr secondary winding; for three-winding xfmrs, $I_{ij}^T$ is through the tertiary winding; for auto-xfmrs, $I_{ij}^S$ is through the series winding and $I_{ij}^C$ is through the common winding.

\vspace{0.1in}
As noted in \cite{lu2017optimal}, most test networks in the literature neglect GSU xfmrs, which are used to connect the output terminals of generators to the transmission network. These xfmrs and the neutral leg ground points they provide are critical when modeling GICs, so they were added to networks that lack of them by using the method discussed in \cite{lu2017optimal}; in this work, it is assumed that GSU xfmrs are delta-gwye xfmrs.

\subsection{Steady-State GMD MLS formulation} \label{subsec:steadystate-mls}

The problem specification for the MLS use case of PMsGMD with second-order cone programming (SOCP) relaxation is introduced below. The SOCP relaxation lifts the product of voltage variables into a high-dimensional space, replacing bilinear voltage product terms with linear terms \cite{Coffrin2015a}. Switching to the full set of ac power flow constraints or relaxations is seamless given the PMs framework.

\vspace{0.1in}
\noindent \textit{1) Objective function:}

\begin{equation}
\label{eq:mls-obj}
\allowdisplaybreaks
\small
min \smashoperator{\sum_{k \in D}}P_k^d z_k^d
\end{equation}
\noindent where $D$ is the set of loads, $P_k^d$ is the scheduled active power demand for load $k$, and $z_k^d \in [0,1]$ is the load-shed variable.

\vspace{0.1in}
\noindent \textit{2) Power flow equations:}

\vspace{0.05in}
\noindent $ \forall i\in {N^a} $
\begin{multline}
\label{eq:mls-kcl}
\allowdisplaybreaks
\small
\sum_{ij \in {E_i^+}} ( \mathbf{S}_{ij} - \mathbf{i} Q_{ij}^{loss} ) - \sum_{ij \in {E_i^-}} \mathbf{S}_{ij} = \\
= \sum_{k \in G_i} z_k^g \mathbf{S}_k^g - \sum_{k \in D_i} z_k^d \mathbf{S}_k^d - \mathbf{Y_i^s}W_{ii}
\end{multline}
\noindent where $N^a$ is the set of ac buses, $E_i^+$ is the set of branches flowing into bus $i$, $E_i^-$ is the set of branches flowing out of bus $i$, $\mathbf{S}_{ij}$ is the complex power flow through branch $ij$, $Q^{loss}_{ij}$ is the reactive power loss of the branch resulting from GIC, $G_i$ is the set of generators connected to bus $i$, $\mathbf{S}_k^g$ is the complex power produced by generator $k$, $z_k^g \in [0,1]$ is the generator status, $D_i$ is the set of loads connected to bus $i$, $\mathbf{S}_k^d$ is the scheduled power of load $k$, $z_k^d \in [0,1]$ is the load shed variable, $\mathbf{Y}_i^s$ is the shunt admittance of bus $i$, and $W_{ii}$ is the voltage at bus $i$ in the lifted space. 

\vspace{0.05in}
\noindent $ \forall i\in {N^a} $
\begin{equation}
\label{eq:mls-qloss}
\allowdisplaybreaks
\small
Q_{ij}^{loss} = K_{ij} \frac{|V_i| I^{eff}_{ij}}{3 S^b_{ij}}
\end{equation}
\noindent where $K_{ij}$ is an non-negative constant that depends on the xfmr construction, $V_i$ is the voltage at bus $i$, and $S_{ij}^b$ is the rated xfmr power; $V_i \approx 1$ is assumed.

\vspace{0.05in}
\noindent $ \forall e \in {E^a} $
\begin{equation}
\label{eq:dc-ots_powflow-approx_2}
\allowdisplaybreaks
\small
\mathbf{S}_{ij} = \left( \mathbf{Y}_{ij}^* - \mathbf{i}\frac{\mathbf{b}_{ij}^c}{2} \right ) \frac{W_{ii}}{|\mathbf{T}_{ij}|^2} - \mathbf{Y}_{ij}^*\frac{W_{ij}^*}{\mathbf{T}_{ij}}
\end{equation}
\noindent where $Y_{ij}$ is the series admittance of branch $ij$, $\mathbf{b}_{ij}^c$ is the line charging capacitance (equal to 0 for xfmrs), $\mathbf{T}_{ij}$ is the turns ratio and phase shift (equal to 1 for lines), and $W_{ij}$ is the product of voltage variables in the lifted space.

\vspace{0.1in}
\noindent \textit{3) Operational limit constraints:}

\vspace{0.05in}
\noindent $ \forall ij \in {E^a} $
\begin{subequations}
\begin{align}
\allowdisplaybreaks
\small
\tan(-\theta_{ij}^\Delta) \le \operatorname{Im}(W_{ij}) \le \tan(\theta_{ij}^\Delta)
\label{eq:soc-angle-cons} \\
|W_{ij}|^2 \le W_{ii}W_{jj}
\label{eq:soc-voltage-cons}
\end{align}
\end{subequations}
\noindent where $\theta_{ij}^\Delta$ is the phase angle limit across branch $ij$.

\vspace{0.05in}
\noindent $ \forall i \in {\mathcal{N}^a} $
\begin{equation}
\allowdisplaybreaks
\small
z_i\underline{v}_i^2 \le W_{ii} \le z_i\overline{v}_i^2 \qquad
W_{ii}^s = \langle z_i^s W_{ii} \rangle
\end{equation}
\noindent where $\underline{v}_i$ and $\overline{v}_i$ are the lower and upper bounds on the voltage at bus $i$, respectively, and $z_i \in [0,1]$ is the bus status variable.

\vspace{0.05in}
\noindent $ \forall k \in \mathcal{G} $
\begin{equation}
\allowdisplaybreaks
\small
\underline{p}_k \leq p^g_k \leq \overline{p}_k \qquad
\underline{q}_k \leq q^g_k \leq \overline{q}_k
\label{eq:mls-gen-power-cons}
\end{equation}
\noindent where $\underline{p}_k^g$ and $\overline{p}_k^g$ are the lower and upper active power limits for generator $k$, while $\underline{q}_k^g$ and $\overline{q}_k^g$ are the reactive power limits.

\subsection{Cascading GMD MLS formulation} \label{subsec:cascading-mls}

The cascading MLS formulation was designed to approximate the quasi-dynamic behavior of a network subjected to GICs. It is intended to model the generator response to area control error (ACE) and continuous load-shedding.

This formulation is identical to the steady-state MLS formulation, but with a few differences.
Generator powers are scaled by the same factor within connected components; these are tied together with a global scaling variable associated with the reference bus.
While not explicitly included in this formulation, generator upper and lower bounds are limited by their ramp rates and the generator setpoints of the previous timestep.
Load powers are scaled by the same factor within connected components; these are tied together with a global scaling variable associated with the reference bus.
Modified constraints can be expressed as follows:

\vspace{0.05in}
\noindent $ \forall i\in {N^a} $
\begin{multline}
\allowdisplaybreaks
\small
\sum_{ij \in {E_i^+}} ( \mathbf{S}_{ij} -\mathbf{i}Q_{ij}^{loss} ) - \sum_{ij \in {E_i^-}} \mathbf{S}_{ij} = \\
= \sum_{k \in G_i} \mathbf{S}_k^g - \sum_{k \in D_i} \mathbf{S}_i^d - \mathbf{Y_i^s}W_{ii}
\label{eq:cascading-mls-kcl}
\end{multline}

\vspace{0.05in}
\noindent $ \forall k\in G $, $ \forall k\in D $
\begin{equation}
\allowdisplaybreaks
\small
\mathbf{S}_k^g = z^g \mathbf{S}_k^{g0} \qquad
\mathbf{S}_k^d = z^d \mathbf{S}_k^{d0}
\label{eq:cascading-mls-scaling}
\end{equation}
\noindent where $z^g$ and $z^d$ are the global generator status and load-shed variables, respectively, while $S_k^{g0}$ and $S_k^{d0}$ are the scheduled generator and load powers determined at the previous iteration; note that buses still retain individual status variables.

\section{Case Study and Results} \label{sec:case-study}
\indent


To demonstrate the use of the GMD Cascade Simulator, a case study was performed: the developed cascading algorithm was evaluated on the created RTS-GMLC-GIC-EAST synthetic test network for the $Scaled$ $A2$ GMD scenario.

The IEEE RTS-GMLC test network \cite{case-rtsgmlc} was extended and modified to enable the time-domain simulation of the impacts of GICs for this study.
Corresponding dc network was created: each resistive branch was replaced with its admittance value and a voltage source in series; transmission lines with series capacitive compensation were omitted as series capacitors block the flow of GIC; after removing by-default disconnected wind and solar generators, individual generator buses were added to each existing generator that connect to the original load buses or substations through added GSU xfmrs.
The network was geolocated to the U.S. East Coast: it was rotated by 130$^{\circ}$ and moved to the region covering Maryland, Virginia, and North Carolina; the three network areas were overlapped with densely populated metropolitan areas to demonstrate GMD risk at those southern latitudes.
The created synthetic network is hereinafter referred to as RTS-GMLC-GIC-EAST.

\begin{figure}[!htbp]
\centering
{
\setlength{\fboxsep}{0pt}
\setlength{\fboxrule}{0.5pt}
\fbox{\includegraphics[width=0.35\textwidth]{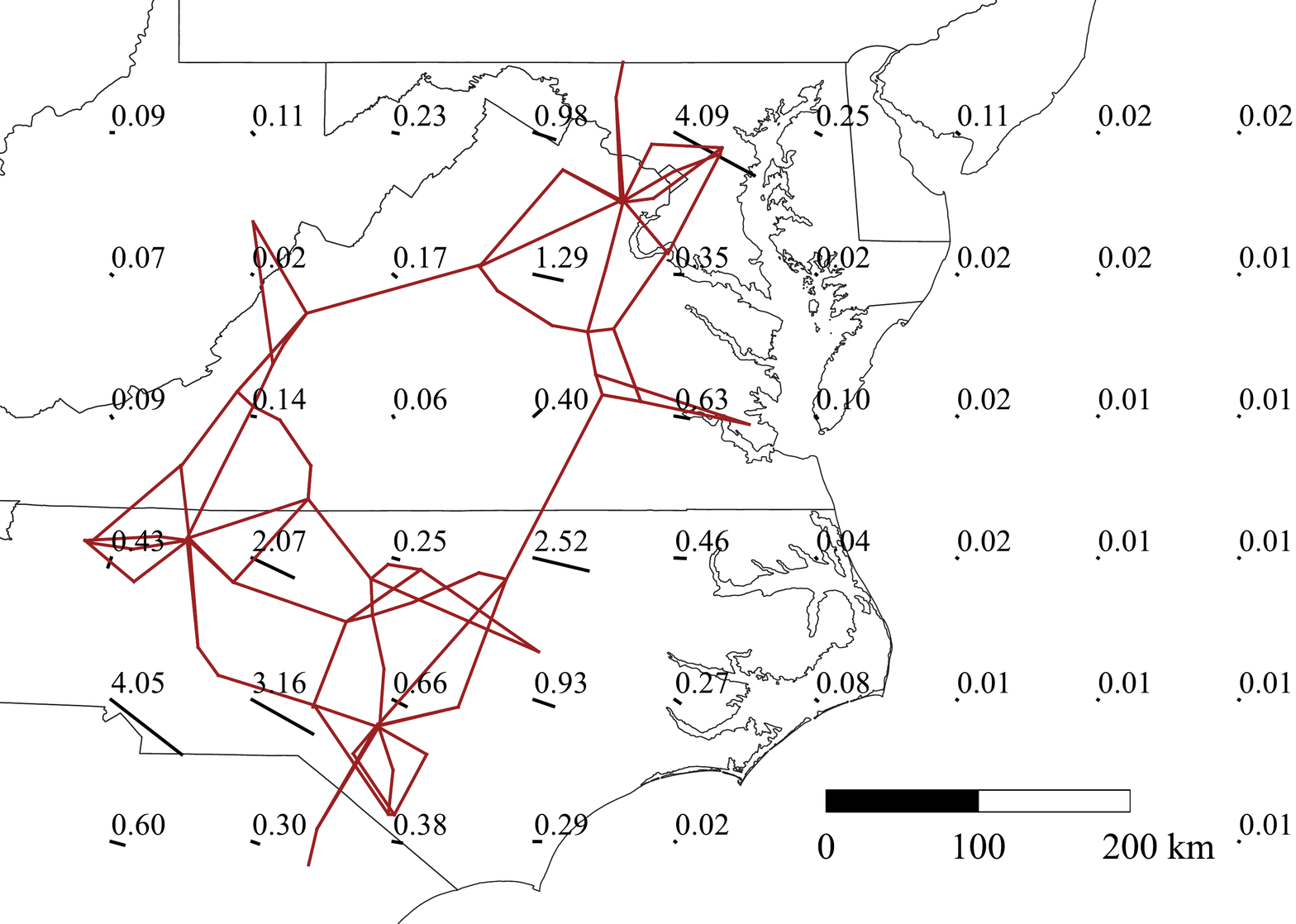}}
}
\caption{
The map of the RTS-GMLC-GIC-EAST synthetic test network, annotated with the geoelectric field values at the time the peak electric field strength was observed. The barbs indicate the direction of the geoelectric field.
}
\label{fig:rts-gmlc-gic-east-e-field}
\end{figure}

RTS-GMLC-GIC-EAST spans an area of approximately 400~km by 600~km, covering the U.S. States of Virginia and South Carolina, with an average line length of 45.5~km and maximum line length of 157.4~km.

\begin{table}[!htbp]
\caption{RTS-GMLC-GIC-EAST Synthetic Test Network}
\begin{center}
\begin{tabular}{lcrl}
\Xhline{2\arrayrulewidth}
Quantity & Value & Units \\ 
\hline 
Number of buses & 169 & -- \\
Number of generators & 96 & -- \\
Number of transmission lines & 105 & -- \\
Number of transformers & 111 & -- \\
Total active power load & 8,550 & MW \\
Total reactive power load & 1,740 & MVar \\
Total available active power generator & 9,076 & MW \\
Total available reactive power generator & 4,406 & MVar \\
\Xhline{2\arrayrulewidth}
\end{tabular}
\end{center}
\label{table:rts-gmlc-gic-summary}
\end{table}

In addition to the GMD scenario, the network was subjected to the following contingencies: 7 generator buses were disabled (Table~\ref{table:initial-bus-outages}); 5 branches were disabled (Table~\ref{table:initial-branch-outages}), causing existing branches to operate near capacity, without losing connected components; and each load was increased by a factor of 1.5. 

\begin{table}[!htbp]
\caption{Initial Bus Outages}
\begin{center}
\begin{tabular}{lcrl}
\Xhline{2\arrayrulewidth}
Number & Name & Base kV \\ 
\hline 
1009 & ALDER\_107\_G & 22 \\
1019 & ASTOR\_118\_G & 22 \\
1052 & BAYLE\_218\_G & 22 \\
1076 & CECIL\_313\_G & 22 \\
1087 & CLARK\_318\_G & 22 \\
1095 & COMPTE\_323\_G1 & 22 \\
1096 & COMPTE\_323\_G2 & 22 \\
\Xhline{2\arrayrulewidth}
\end{tabular}
\end{center}
\label{table:initial-bus-outages}
\end{table}

\begin{table}[!htbp]
\caption{Initial Branch Outages}
\begin{center}
\begin{tabular}{lcrl}
\Xhline{2\arrayrulewidth}
From Bus Number & To Bus Number & Circuit ID \\ 
\hline 
104 & 109 & ``1 '' \\
101 & 103 & ``1 '' \\
123 & 217 & ``1 '' \\
113 & 215 & ``1 '' \\
105 & 110 & ``1 '' \\
\Xhline{2\arrayrulewidth}
\end{tabular}
\end{center}
\label{table:initial-branch-outages}
\end{table}





\begin{figure*}[!htbp]
\centering
\subfloat[\textit{Average coupled line voltage.}]{\includegraphics[width=2.5in]{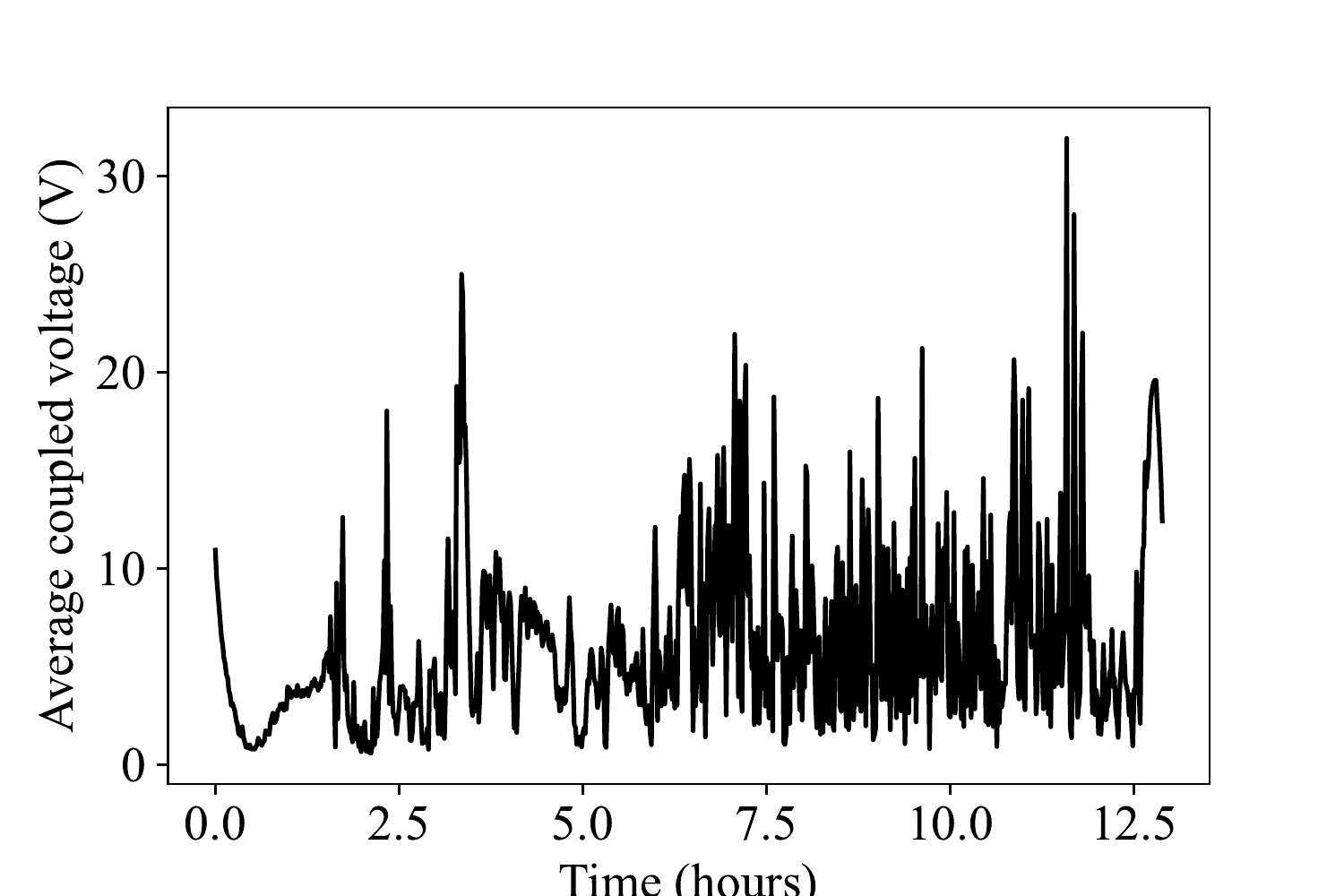}
\label{fig:avg-coupled-voltage}}
\hfil
\subfloat[\textit{Maximum coupled line voltage.}]{\includegraphics[width=2.5in]{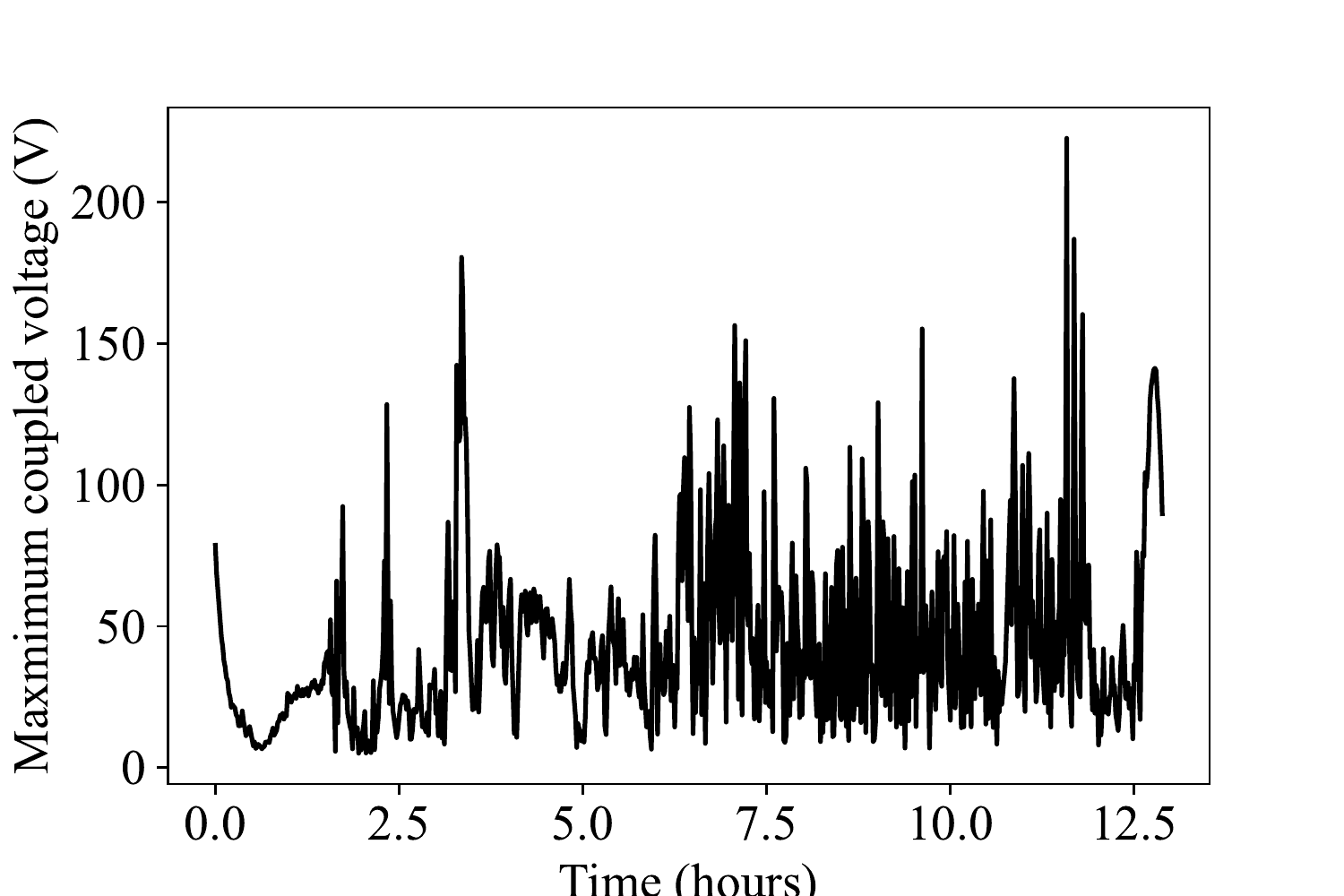}
\label{fig:max-coupled-voltage}}
\hfil
\subfloat[\textit{Branch statuses.}]{\includegraphics[width=2.5in]{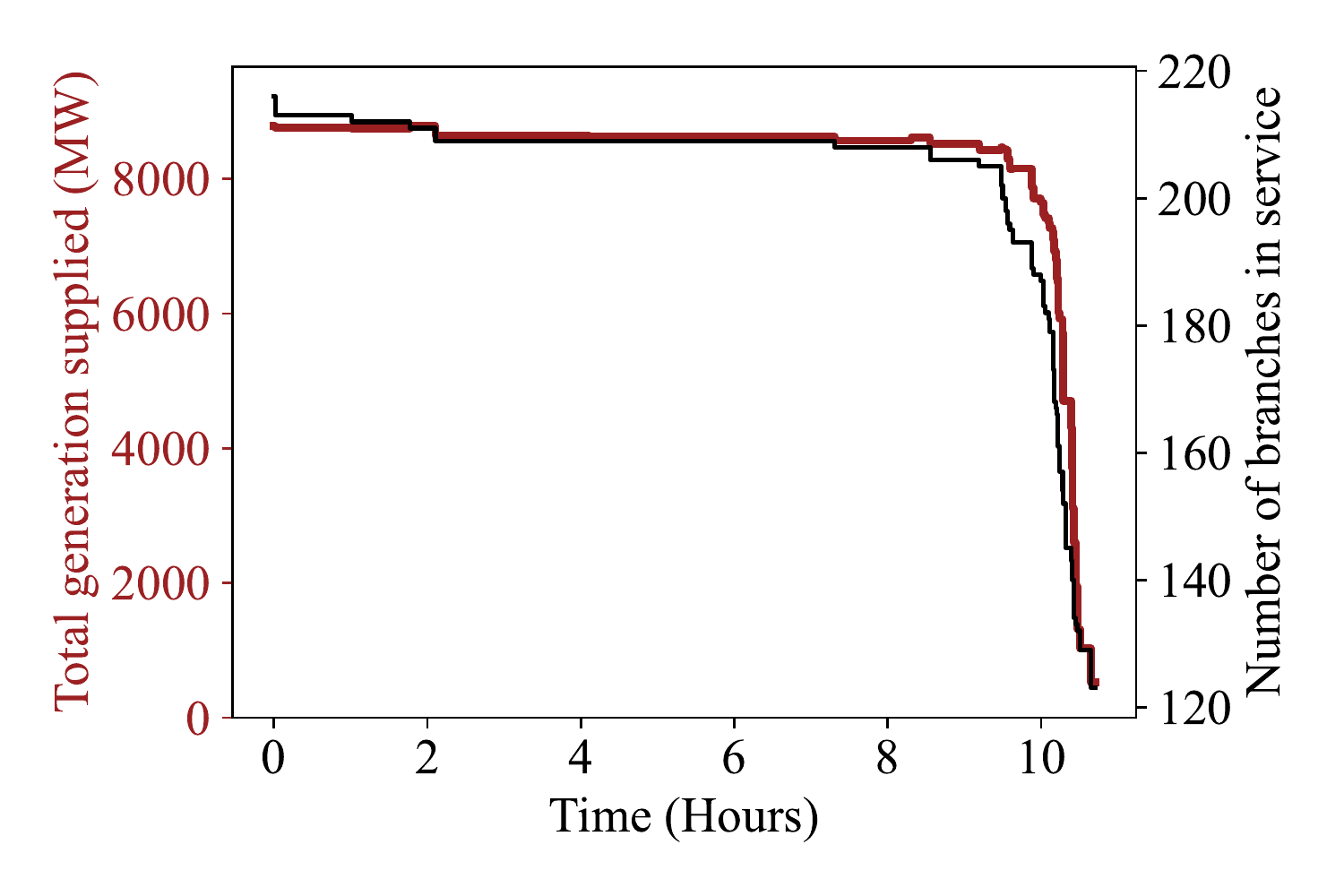}
\label{fig:branch-status}}
\caption{Simulation results for the RTS-GMLC-GIC-EAST synthetic test network during the $Scaled$ $A2$ scenario \cite{blake2021sims}.}
\label{fig:simulation-results}
\end{figure*}

Figure~\ref{fig:simulation-results} (on page \pageref{fig:simulation-results}) presents the results of the performed case study: (a) shows the average coupled line voltage, (b) shows the maximum coupled line voltage, and (c) shows the total generation and total number of online branches during the 12.5~hour simulation period of the analyzed GMD scenario.
Simulations were done using version\_0.4 of PMsGMD, and the computation time is approximately 1.43~sec per iteration for the 169 bus case study system.
Throughout the entire simulation, the total generator active power remained constant 6,065~MW, which means that maximum power was generated and delivered in the network up until a sufficient number of branches become disabled, at which point the network quickly transitions to a blackout state.

\vspace{0.1in}
As the threat of GMD events and caused cascading outages continues to pose a substantial -- foreseeably growing -- risk to the United States electrical grid, accurate simulation and effective mitigation becomes increasingly important.
The presented PMsGMD based GMD Cascade Simulator is a practical tool for modeling and analyzing the impacts of such hazards.

By leveraging the flexible and extensible open-source PMsGMD software, a completely novel, specific GMD-induced cascading failure modeling implementation was developed.
The case study showcased the ability to approximate the dynamic behavior of generators (e.g., operating under ACE, generator relay tripping, load under-frequency relay tripping) in cascading simulation with a relaxed MLS formulation that provides guaranteed convergence; the cascading algorithm results are highly correlated to fully dynamic modeling results, but extensive data requirements are eliminated.
The adaptation of DCSIMSEP and its customization via incorporated ac relaxations allows for realistic simulation of cascading outages and enables to analyze failures in various initiating locations that are dependent on the particular GMD scenario conditions.
As commercial software is currently without similar ability, this new tool further increases the value of PMsGMD for both academic research and industry use.


\bibliographystyle{unsrt}
\bibliography{references}

\end{document}